\documentclass[pra,aps,twocolumn,superscriptaddress]{revtex4}
\usepackage{hyperref}
\usepackage{bm}
\usepackage{epsfig}
\usepackage{amssymb}
\usepackage{graphicx}
\usepackage{epstopdf}
\usepackage{bbold}

\usepackage{amsmath}
\usepackage{dsfont}

\begin{document}
\title{Entangling the whole by beam splitting a part}
\author{Callum Croal$^*$}
\affiliation{School of Physics and Astronomy, University of St.
Andrews, North Haugh, St. Andrews, Fife, KY16 9SS, Scotland}
\author{Christian Peuntinger$^*$}
\affiliation{Max Planck Institute for the Science of Light,
G\"unther-Scharowsky-Str. 1/Bldg. 24, Erlangen, Germany}
\affiliation{Institute of Optics, Information and Photonics,
University of Erlangen-Nuremberg, Staudtstra{\ss}e 7/B2, Erlangen,
Germany}
\author{Vanessa Chille}
\affiliation{Max Planck Institute for the Science of Light,
G\"unther-Scharowsky-Str. 1/Bldg. 24, Erlangen, Germany}
\affiliation{Institute of Optics, Information and Photonics,
University of Erlangen-Nuremberg, Staudtstra{\ss}e 7/B2, Erlangen,
Germany}
\author{Christoph Marquardt}
\affiliation{Max Planck Institute for the Science of Light,
G\"unther-Scharowsky-Str. 1/Bldg. 24, Erlangen, Germany}
\affiliation{Institute of Optics, Information and Photonics,
University of Erlangen-Nuremberg, Staudtstra{\ss}e 7/B2, Erlangen,
Germany}
\author{Gerd Leuchs}
\affiliation{Max Planck Institute for the Science of Light,
G\"unther-Scharowsky-Str. 1/Bldg. 24, Erlangen, Germany}
\affiliation{Institute of Optics, Information and Photonics,
University of Erlangen-Nuremberg, Staudtstra{\ss}e 7/B2, Erlangen,
Germany}
\author{Natalia Korolkova}
\affiliation{School of Physics and Astronomy, University of St.
Andrews, North Haugh, St. Andrews, Fife, KY16 9SS, Scotland}
\author{Ladislav Mi\v{s}ta, Jr.}
\affiliation{Department of Optics, Palack\' y University, 17.
listopadu 12,  771~46 Olomouc, Czech Republic}

\date{\today}

\begin{abstract}
A beam splitter is a basic linear optical element appearing in
many optics experiments and is frequently used as a
continuous-variable entangler transforming a pair of input modes
from a separable Gaussian state into an entangled state. However,
a beam splitter is a passive operation that can create
entanglement from Gaussian states only under certain conditions.
One such condition is that the input light is suitably squeezed.
We demonstrate experimentally that a beam splitter can create
entanglement even from modes which do not possess such a squeezing
provided that they are correlated to but not entangled with a
third mode. Specifically, we show that a beam splitter can create
three-mode entanglement by acting on two modes of a three-mode
fully separable Gaussian state without entangling the two modes
themselves. This beam splitter property is a key mechanism behind
the performance of the protocol for entanglement distribution by
separable states. Moreover, the property also finds application in
collaborative quantum dense coding in which decoding of
transmitted information is assisted by interference with a mode of
the collaborating party.

\end{abstract}

\maketitle


A beam splitter (BS) is an optical device that can superimpose
incident light modes. As quadrature amplitudes of the incoming
modes are also superimposed in this process, correlations may
arise between the corresponding quadratures of the output modes.
In particular, if modes squeezed in conjugate quadratures enter a
BS, 
an entangled state carrying Einstein-Podolsky-Rosen correlations
\cite{Einstein_35} emerges at the output
\cite{Furusawa_98,Silberhorn_01}. This is currently a widely used
experimental method, which finds application 
in
continuous-variable (CV) quantum teleportation \cite{Furusawa_98},
dense coding \cite{Li_02} or cryptography
\cite{Ralph_00,Silberhorn_02}.

Whether entanglement will be created by a BS depends on the nature
of the input states. For classical input states given by
statistical mixtures of coherent states the output states are also
classical \cite{Xiang-bin_02} and thus possess no entanglement.
Therefore, to get entanglement on a BS some non-classicality is
needed at the input \cite{Kim_02,Paris_15}. In CV experiments
non-classicality of Gaussian states
\cite{Braunstein_05+Weedbrook_12} is used for this purpose which
is equivalent to squeezing \cite{Fiurasek_review01}. However, the
mere presence of some input squeezing may not suffice for the
generation of entanglement on a BS \cite{Kim_02}.  It requires
more stringent conditions \cite{Wolf_03}. An interesting question
that arises is whether interference on a BS of some states which
do not satisfy the condition and hence do not entangle on a BS can
still create some entanglement. Remarkably, this is indeed
possible if the interfered state is a local state of a fully
separable state of a larger system. This is illustrated by the
protocol for Gaussian entanglement distribution by separable
states \cite{Mista_08+09} where a BS creates entanglement by
acting on two modes of a fully separable three-mode Gaussian state
while leaving the modes individually disentangled. Note, that so
far such a property of a BS has not been demonstrated, because in
the experiment \cite{Peuntinger_13} the additional third mode was
independent of the superimposed modes, whereas in the experiment
\cite{Vollmer_13}, the third mode was entangled with one of the modes
entering the BS.

In this Letter we provide an experimental demonstration of the
property of a BS mentioned above using two examples. In the
first example the initial separable state is two-mode and it is
prepared by random displacements of a squeezed state and a vacuum
state in one quadrature. By splitting the former state on a BS we
then create a three-mode state in which the output modes of the BS are
separable individually but simultaneously each output mode is
entangled with the remaining two modes. In the second example we
prepare a three-mode fully separable state by random displacement
of two orthogonally squeezed states in the squeezed quadratures
and the vacuum state in both quadratures. By superimposing
originally squeezed states on a BS we then create an entangled
state in which the output modes of the BS are again separable but
one of the output modes is entangled with the remaining two modes
at the same time. The state from the first (second) example is a
backbone of the CV protocol for entanglement sharing
\cite{Mista_13} (distribution \cite{Mista_08+09}) with separable
states. Moreover, despite being partially separable and noisy, the
states also enable assisted quantum dense coding which can beat
coherent-state and even squeezed-state communication capacity.


We demonstrate the aforementioned effect using quantum modes of the
electromagnetic field, which are quantum systems with
infinite-dimensional Hilbert spaces. A system of $n$ modes
is described by the vector of quadratures
$\hat{\boldsymbol{\xi}}=(\hat{x}_1,\hat{p}_1,...,\hat{x}_n,\hat{p}_n)$,
the elements of which satisfy the canonical commutation rules
$[\hat{\xi}_j,\hat{\xi}_k]=i(\boldsymbol{\Omega}_{n})_{jk}$, with
$\boldsymbol{\Omega}_n=\oplus_{j=1}^{n}i\boldsymbol{\sigma}_{y}$,
where $\boldsymbol{\sigma}_{y}$ is the Pauli-$y$ matrix. Quantum
states involved in our experiment are well approximated by
Gaussian states \cite{Peuntinger_13}, i.e., states with a Gaussian
Wigner function, and we resort to the tools of Gaussian
quantum information theory \cite{Braunstein_05+Weedbrook_12} in
what follows.

A Gaussian state $\hat{\rho}$ is fully characterized by the vector
of first moments
$\langle\hat{\boldsymbol{\xi}}\rangle\equiv\text{Tr}(\hat{\boldsymbol{\xi}}\hat{\rho})$,
which is always zero in the present case, and the covariance
matrix (CM) $\boldsymbol{\gamma}$ with elements
$\gamma_{jk}=\langle\hat{\xi}_{j}\hat{\xi}_k+\hat{\xi}_k\hat{\xi}_j\rangle-2\langle\hat{\xi}_j\rangle\langle\hat{\xi}_k\rangle$.
An important property of a Gaussian state is its non-classicality.
We say that a state is non-classical if it cannot be expressed as
a statistical mixture of coherent states
\cite{Glauber_63,Sudarshan_63}. In the Gaussian scenario, non-classicality
is equivalent with squeezing \cite{Fiurasek_review01}, i.e., a
Gaussian state with CM $\boldsymbol{\gamma}$ is non-classical if
and only if (iff)
$\mu:=\mbox{min}[\mbox{eig}(\boldsymbol{\gamma})]<1$
\cite{Simon_94}.

In the present experiment we generate three-mode states with
specific separability properties. We certify the properties using
the positive partial transpose (PPT) criterion for Gaussian states
\cite{Werner_01,Duan_00,Simon_00}. The partial transposition
operation with respect to mode $j$ transforms an $n$-mode CM
$\boldsymbol{\gamma}$ to
$\boldsymbol{\gamma}^{(T_j)}=\boldsymbol{\Lambda}_j\boldsymbol{\gamma}\boldsymbol{\Lambda}_j$
with $\boldsymbol{\Lambda}_j=(\oplus_{i\ne
j=1}^{n}\mathbb{1}^{(i)})\oplus\boldsymbol{\sigma}_z^{(j)}$, where
$\mathbb{1}^{(i)}$ is the 2$\times$2 identity matrix of mode $i$,
and $\boldsymbol{\sigma}_z^{(j)}$ is the Pauli-$z$ matrix of mode
$j$. The PPT criterion then states that the state is separable
with respect to mode $j$ iff
\begin{equation}\label{separability}
\boldsymbol{\gamma}^{(T_j)}+i\boldsymbol{\Omega}_n\ge0.
\end{equation}


\begin{figure}[tb]
\vspace{5mm}
\includegraphics[width=7.5cm,angle=0]{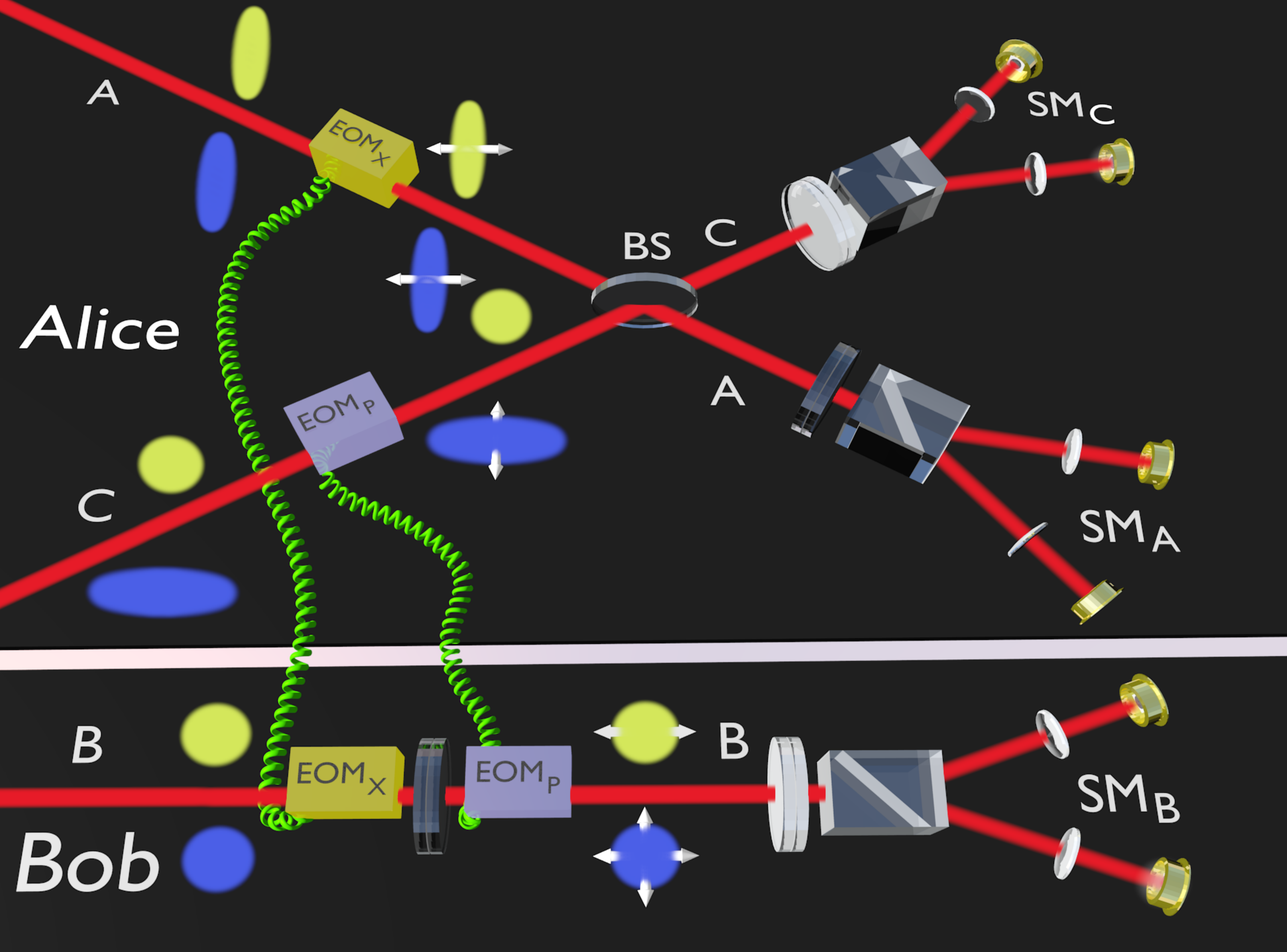}
\caption{Experimental scheme. $\mbox{EOM}_i$: electro-optical
modulator displacing quadrature $i$, BS: balanced beam splitter,
$\mbox{SM}_j$: Stokes measurement on mode $j$. The yellow (blue)
circles and ellipses represent the states of protocol 1 (2). The
yellow modulators $\mbox{EOM}_x$ are applied to both protocols and
the blue modulators $\mbox{EOM}_p$ are only applied to protocol
2.}\label{fig1}
\end{figure}

We demonstrate the entangling power of a BS on two protocols
depicted in Fig.~\ref{fig1}. First, we focus on a more simple
protocol (yellow circles and ellipses in Fig.~\ref{fig1}) which we
shall refer to as protocol 1.  Alice initially holds mode $A$
in a position squeezed vacuum state with quadratures $\hat
x_{A}=e^{-r}\hat x_{A}^{(0)}$, $\hat p_{A}=e^{r}\hat p_{A}^{(0)}$,
where $r>0$ is the squeezing parameter and the superscript
``$(0)$'' denotes the vacuum quadratures, while Bob holds a vacuum
mode $B$ with quadratures $\hat x_{B}=\hat x_{B}^{(0)}$ and $\hat
p_{B}=\hat p_{B}^{(0)}$. Next, modes $A$ and $B$ are displaced as
\begin{equation}
\hat{x}_A\rightarrow\hat{x}_A+x,\quad\hat{x}_B\rightarrow\hat{x}_B+x,
\end{equation}
where the classical displacement $x$ is Gaussian distributed with
zero mean and variance $\langle x^2\rangle=V_{1}:=(1-e^{-2r})/2$.
The displacements realize an LOCC operation and the resulting
state is thus separable. Further, the displacements are strong
enough to destroy the initial squeezing and the state of mode $A$
is classical. However, the lowest eigenvalue of the CM of the
global state is equal to $\mu=e^{-r}[\cosh r-(\sqrt{5}-2)\sinh
r]<1$ and hence the state is non-classical.

Mode $A$ is then split on a BS realizing a transformation
$\hat{x}_{A,C}\rightarrow(\hat{x}_{A}\pm\hat{x}_{C})/\sqrt{2},\quad
\hat{p}_{A,C}\rightarrow(\hat{p}_{A}\pm\hat{p}_{C})/\sqrt{2}$,
where $\hat x_{C}=\hat x_{C}^{(0)}$ and $\hat p_{C}=\hat
p_{C}^{(0)}$ are quadratures of the vacuum mode $C$ entering the
empty port of the BS. Application of the BS results in a
three-mode state with CM $\boldsymbol{\gamma}_{1}$, which carries
no entanglement between any two modes and across the $B|AC$ splitting,
but which is entangled across the $A|BC$ and $C|AB$ splittings
\cite{Mista_13}. The protocol 1 thus demonstrates the required
entangling property of a BS. Indeed, we have created entanglement
from a fully separable three-mode Gaussian state by mixing on a BS
two modes of the state while these two modes alone remain separable after the BS.

The entangling capability of a BS can be also illustrated using a
different protocol called protocol 2 in what follows (blue circles
and ellipses in Fig.~\ref{fig1}).
Alice holds mode $A$ in a position-squeezed state and mode $C$ in
a momentum squeezed state, and Bob holds mode $B$ in a vacuum
state. The modes are then displaced as
\begin{eqnarray}
\hat{x}_A&\rightarrow&\hat{x}_A+x,\quad\hat{x}_B\rightarrow\hat{x}_B+\sqrt{2}x,\nonumber\\
\hat{p}_C&\rightarrow&\hat{p}_C-p,\quad\hat{p}_B\rightarrow\hat{p}_B+\sqrt{2}p,
\end{eqnarray}
where $x$ and $p$ are uncorrelated classical displacements obeying
zero mean Gaussian distributions with variances $\langle
x^2\rangle=\langle p^2\rangle=V_{2}:=(e^{2r}-1)/2$. In comparison
with protocol 1, the final state with CM $\boldsymbol{\gamma}_{2}$
of protocol 2 has different separability properties. Specifically,
the state is separable across the $B|AC$ and $C|AB$ splittings,
which guarantees absence of any two-mode entanglement, and it is
entangled across the $A|BC$ splitting. Thus again, entanglement is
created by mixing on a BS two modes of a three-mode fully
separable Gaussian state, whereas the two modes at the output of
the BS do not get entangled.

The experimental setup for both protocols is shown in
Fig.~\ref{fig1}. The implementation is realized using Stokes
observables and measurements. A high excitation of $\hat{S}_3$
(circular polarization), and in contrast
$\langle\hat{S}_1\rangle=\langle\hat{S}_2\rangle = 0$, allows the
$S_1$-$S_2$-plane (also called ``dark-plane''\cite{Heersink_05}) to be interpreted
as the quadrature phase space. Within this plane $\hat{S_\theta}$
($\hat{S}_{\theta+{\pi/2}}$) is identified with the $\hat{x}$
($\hat{p}$) quadrature. In both scenarios Alice possesses two
modes $A$ and $C$ and Bob holds mode $B$. All the involved modes
(except mode $C$ in the first protocol, which is a real vacuum
mode) are created with a soliton laser (Origami, Onefive, center
wavelength: 1559\,nm, pulse length: 200\,fs, repetition rate:
80\,MHz).
Mode $A$ initially exhibits polarization squeezing in
$\hat{S}_{\theta}$ ($\hat{x}$), generated by using the Kerr
nonlinearity of an optical fiber (FS-PM-7811, Thorlabs, 13
m)~\cite{Heersink_05, Leuchs_99,Silberhorn_01, Dong_07}. This
initial squeezing is destroyed by modulation in the squeezing
direction $\hat{S}_\theta$ realized in two steps. First, different
displacements are realized by slightly modulating the state of
polarization applying a sinusoidal voltage (frequency: 18.2\,MHz)
to an electro optical modulator (EOM) and electronically
down-mixing the Stokes measurement signal with a phase matched
local oscillator of the same frequency. Second, digitally mixing
measurements from different displaced modes leads to a mixed
Gaussian state, which can be expressed as a statistical mixture of
coherent states. In the simpler protocol this mode $A$ is mixed
with a vacuum mode $C$ on a BS. Bob prepares mode $B$, initially
coherent but modulated in the same manner and correlated to mode
$A$ of Alice
. The CM $\boldsymbol{\gamma}_{1}$ was
measured to be
\begin{eqnarray*}\label{gamma1}
\boldsymbol{\gamma}_1=\left(\begin{array}{cccccc}
5.42 & 0.23 & 3.34 & -0.73 & 4.06 & 0.04\\
0.23 & 19.28 & 0.00 & 0.00 & 0.45 & 17.29\\
3.34 & 0.00 & 3.43 & -0.54 & 3.06 & -0.03\\
-0.73 & 0.00 & -0.54 & 1.12 & -0.67 & 0.01 \\
4.06 & 0.45 & 3.06 & -0.67 & 4.73 & 0.55\\
0.04 & 17.29 & -0.03 & 0.01 & 0.55 & 17.70\\
\end{array}\right).
\end{eqnarray*}
The measurement errors for the elements of all measured CMs lie between $0.002$ and $0.023$.

As for protocol 2, it uses the same mode $A$ as the first one,
but it is extended with a mode $C$. This mode $C$  is prepared in the same way as mode $A$,
just the squeezing and, accordingly, also the modulation are in
the $\hat{S}_{\theta+{\pi/2}}$ direction. Again
these two modes are mixed on a BS. In this case, Bob's mode $B$ is
independently modulated in both conjugate Stokes observables
$\hat{S}_\theta$ and  $\hat{S}_{\theta+{\pi/2}}$. In the case of
protocol 2, the CM $\boldsymbol{\gamma}_{2}$ was measured to be
\begin{eqnarray*}\label{gamma2}
\boldsymbol{\gamma}_2=\left(\begin{array}{cccccc}
20.90 & 1.10 & 5.17 & -8.59 & -7.80 & -1.68\\
1.10 & 25.31 & -5.04 & -6.76 & 1.00 & 14.64\\
5.17 & -5.04 & 11.87 & -0.45 & 4.95 & 4.49\\
-8.59 & -6.76 & -0.45 & 18.88 & -8.61 & 6.04\\
-7.80 & 1.00 & 4.95 & -8.61 & 20.68 & 0.80\\
-1.68 & 14.64 & 4.49 & 6.04 & 0.80 & 24.65\\
\end{array}\right).
\end{eqnarray*}

The architecture of our experimental setup together with the
separability properties of the measured CMs guarantee that we
are able to really observe the predicted entangling capability of a BS.
Note first, that the three-mode state before the BS has been
prepared by local operations on independent modes and classical
communication (green wires in Fig.~\ref{fig1}) and therefore it is
by construction fully separable. Further, by applying the
separability criterion (\ref{separability}) on CMs
$\boldsymbol{\gamma}_{1}$ and $\boldsymbol{\gamma}_{2}$ and the
local CMs $\boldsymbol{\gamma}_{1,AC}$ and
$\boldsymbol{\gamma}_{2,AC}$ of reduced states of modes $A$ and
$C$, we can confirm that the CMs also exhibit the desired
separability properties.

The three-mode separability properties of the CMs
$\boldsymbol{\gamma}_{1}$ and $\boldsymbol{\gamma}_{2}$ are
summarized in Table~\ref{sepeigenvalues}.
\begin{table}[ht]
\caption{Minimum eigenvalue
$\lambda_{k}^{(T_j)}:=\mbox{min}[\mbox{eig}(\boldsymbol{\gamma}_{k}^{(T_j)}+i\boldsymbol{\Omega}_3)]$.}
\centering \label{sepeigenvalues}
\begin{tabular}{| c | c | c | c |}
\hline $j$ & A & B & C \\
\hline $\lambda_{1}^{(T_j)}\times10^{2}$ & $-2.2\pm0.1$ & $6.9\pm0.1$ & $-2.2\pm0.1$  \\
\hline $\lambda_{2}^{(T_j)}\times10$ & $-1.44\pm0.01$ & $3.51\pm0.02$ & $5.28\pm0.03$\\
\hline
\end{tabular}
\end{table}
For CM $\boldsymbol{\gamma}_1$, the two negative minimum eigenvalues in the table reveal that the BS
created entanglement with respect to the $A|BC$ and $C|AB$
splittings, whereas the state is separable across the $B|AC$
splitting, as predicted by the theory. However, as required, at the same time modes $A$ and $C$ did not get entangled according to the criterion (\ref{separability}). This is
evidenced by $\mbox{min}[\mbox{eig}(\boldsymbol{\gamma}_{1,AC}^{(T_A)}+i\boldsymbol{\Omega}_2)]=0.84\pm0.01>0$.

Moving to the CM $\boldsymbol{\gamma}_{2}$, one can see from
Table~\ref{sepeigenvalues} that the CM represents an entangled
state across the $A|BC$ splitting whereas it exhibits separability
across the $B|AC$ and $C|AB$ splittings in accordance with the
theory. Finally, since $\mbox{min}[\mbox{eig}(\boldsymbol{\gamma}_{2,AC}^{(T_A)}+i\boldsymbol{\Omega}_2)]=9.371\pm0.005>0$
\cite{Peuntinger_13}, modes $A$ and $C$ are separable as expected.

The present experiment demonstrates that a BS can create
entanglement even by mixing two modes, which alone cannot be
entangled by the BS. The condition under which this can happen is that the two modes are  part of a
three-mode fully separable system. The entanglement is created
solely by the BS because no entanglement is present before the BS.
The entanglement does not occur between the output modes of the BS
but instead it emerges between one output mode and the remaining
two modes taken together. This phenomenon is a key element of the
protocols for entanglement sharing \cite{Mista_13} and
distribution \cite{Cubitt_03,Mista_08+09} with separable states.
The schemes depicted in Fig.~\ref{fig1} are effectively the first
two steps of these protocols.
Thus a passive BS operation on a tailored three-mode fully
separable state not only can generate entanglement across some
bipartite splittings of a global state, but a further BS can
localize this entanglement between modes $A$ and $B$.
This has been experimentally demonstrated for the entanglement
sharing protocol in \cite{Chille_14} and for the entanglement
distribution protocol in \cite{Peuntinger_13}. In both cases, the
recovery of two-mode entanglement has been performed
electronically on the outcomes of the measurement on mode $C$ and
the presence of entanglement has been certified by the sufficient
condition for entanglement \cite{Giovannetti_02}.


\begin{figure}[tb]
\includegraphics[width=8.0cm,angle=0]{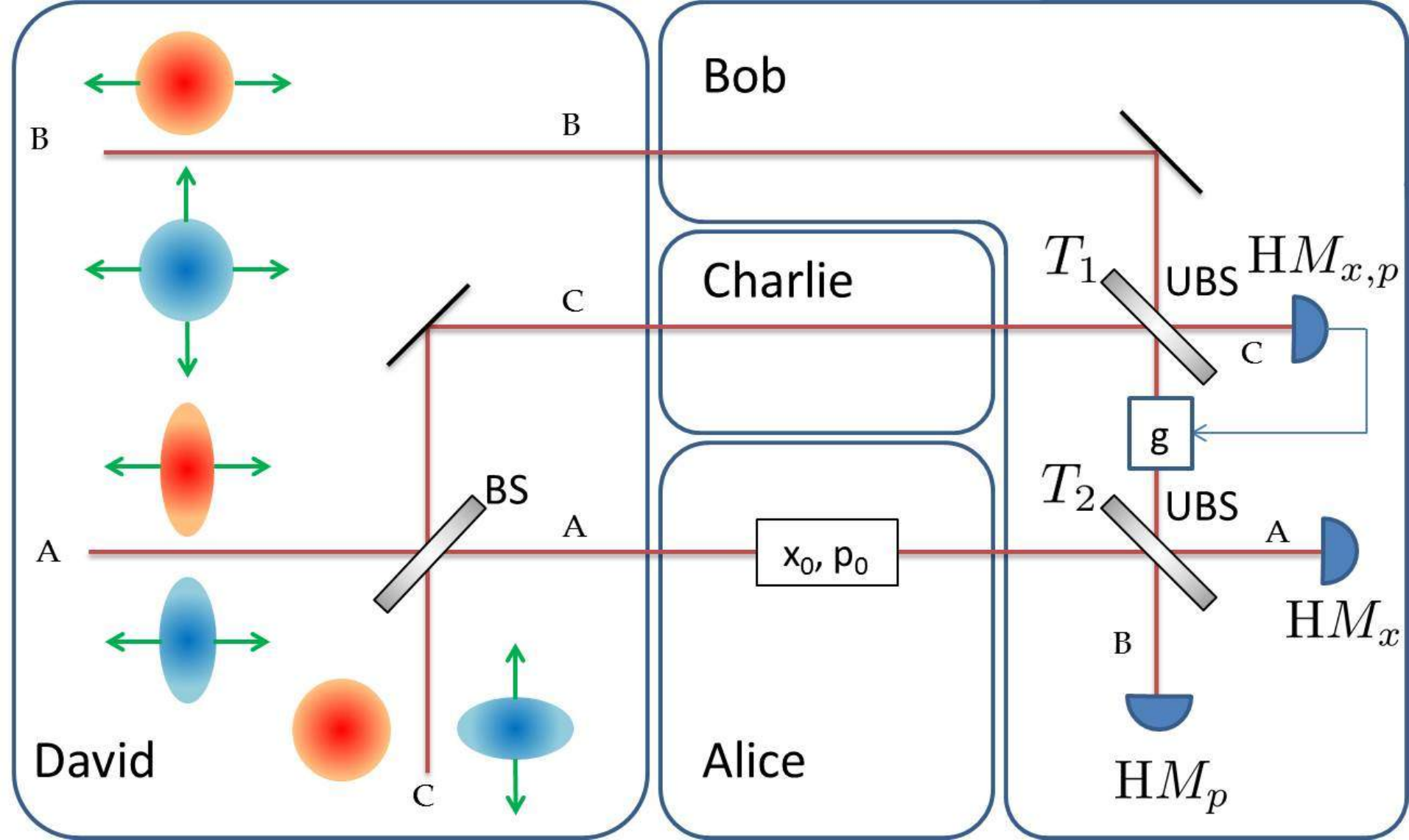}
\caption{Collaborative dense coding schemes with a state from
protocol 1 (orange circles and ellipsis) and protocol 2 (blue
circle and ellipses). BS: balanced beam splitter, UBS: unbalanced
beam splitters with transmissivities $T_1$ and $T_2$, $\mbox{HM}$:
homodyne measurement.}
\label{fig2}
\end{figure}

Entanglement created in protocols 1 and 2 is not only useful for
sharing and distribution of entanglement, but also directly finds
an application in a collaborative version of quantum dense coding
\cite{Bennett_92} with continuous variables
\cite{Braunstein_00,Li_02}. The corresponding scheme is depicted
in Fig.~\ref{fig2}. Comparing to the standard dense coding schemes
containing only a sender Alice and a receiver Bob, in the
collaborative schemes, Charlie controls the capacity of
information transmission between Alice and Bob. While previous
collaborative schemes \cite{Hao_01,Li_02} were based on genuine
tripartite entanglement and the control of capacity was
accomplished by a measurement on Charlie's mode, the present
scheme relies only on a partially entangled tripartite state and
it utilizes interference of the collaborator and receiver's mode
for the control.

The schemes in Fig.~\ref{fig2} start with preparation of the
output state of protocol 1 (2) about which the participants have
no information. To emphasize this, we attribute the preparation to
a separate party, David. After running protocol 1 (2) David
distributes modes $A,B$ and $C$ of the state with CM
$\boldsymbol{\gamma}_{1}$ ($\boldsymbol{\gamma}_{2}$) to Alice,
Bob and Charlie. Alice encodes on her mode classical Gaussian
signals $x_{0}$ and $p_{0}$ with variance $P$ by performing
displacements $\hat{x}_{A}\rightarrow \hat{x}_{A}+x_{0}$ and
$\hat{p}_{A}\rightarrow \hat{p}_{A}+p_{0}$ and sends the mode to Bob. Upon receiving the mode, Bob decodes
the signal with the help of Charlie in two steps depicted in
Fig.~\ref{fig2}. First, he superimposes his mode with mode $C$ on
an unbalanced BS $\hat{\alpha}_{B,C}'=R_{1}\hat{\alpha}_{C,B}\pm
T_{1}\hat{\alpha}_{B,C}$, $\alpha=x,p$, measures the quadrature
$\hat{p}_{C}'$ on output mode $C$ with outcome $\bar{p}$, and
displaces the other output mode $B$ as
$\hat{p}_{B}'\rightarrow\hat{p}_{B}'+g\bar{p}$, where a gain $g$
maximizes the capacity. In the second step, Bob superimposes modes
$A$ and $B$ on another unbalanced BS with transmissivity $T_{2}$
and measures the quadrature $\hat{x}$ ($\hat{p}$) on output mode
$A$ ($B$). Making use of the formula for capacity of a
communication channel with Gaussian distributed signal (noise) of
power $S$ ($N$), $C=(1/2)\ln(1+S/N)$ \cite{Shannon_48}, we have
then calculated the channel capacity (CC) $C^{j}$ for the protocol
$j$.

\begin{figure}[b]
\includegraphics[clip=false,width=7.0cm,angle=0]{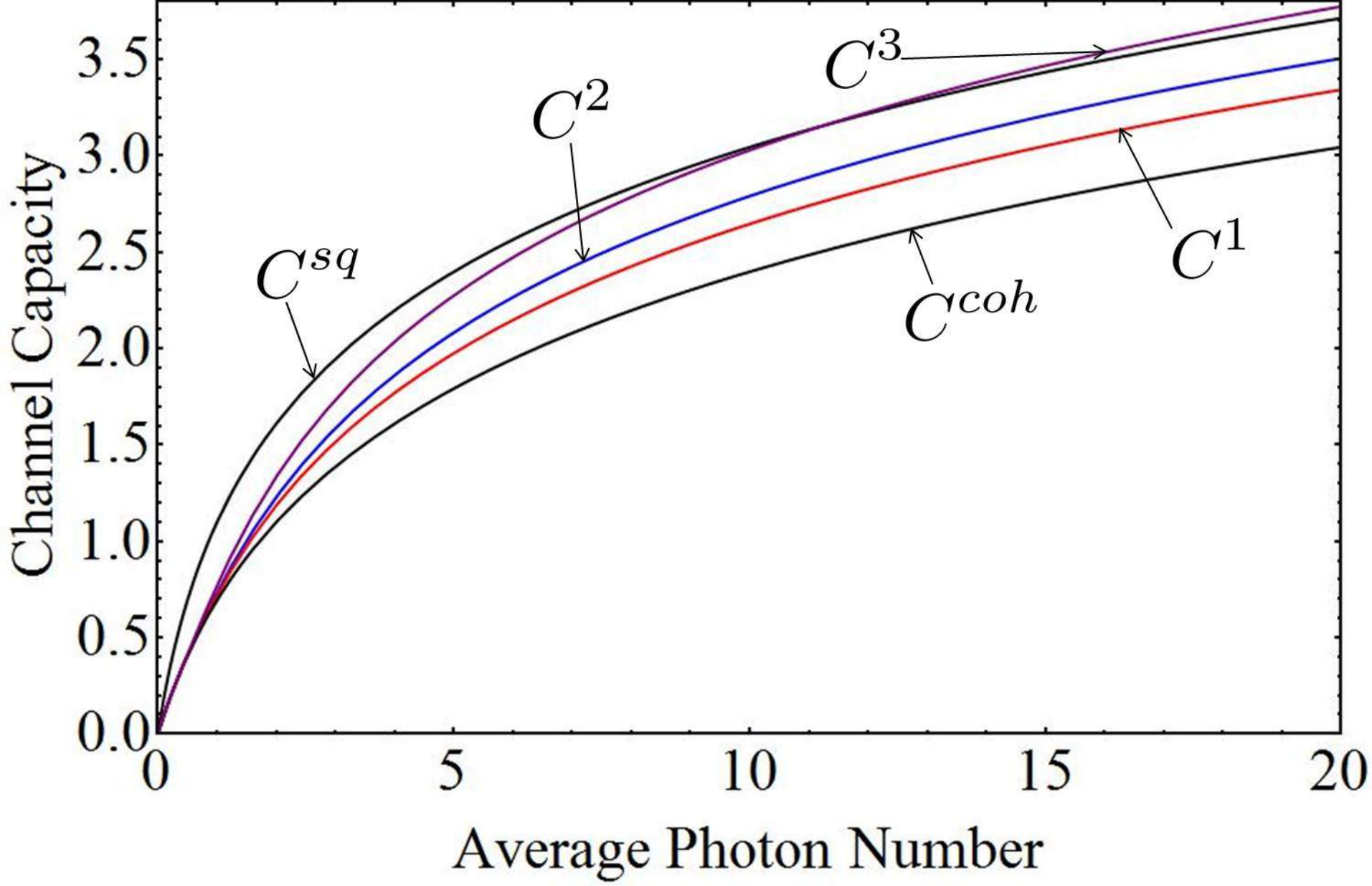}
\caption{Channel capacities $C^{1}, C^{2}$ and $C^{3}$ for the
protocols 1,2 and 3 versus the average photon number. $C^{\rm sq}$
($C^{\rm coh}$): capacity of squeezed (coherent) state
communication.}\label{fig3}
\end{figure}

The CC has been optimized over the transmissivities $T_{1}$ and
$T_{2}$ for fixed average photon number $\bar{n}$ and it is
plotted in Fig.~\ref{fig3}. For comparison, we plot
also the capacities for coherent-state communication with
heterodyne detection, $C^{\rm coh}=\ln(1+\bar{n})$, and the
squeezed-state communication with homodyne detection, $C^{\rm
sq}=\ln(1+2\bar{n})$ \cite{Yamamoto_86}. Besides, we have also
considered a third protocol 3 with the CC $C^3$, which is the same
as protocol 2, but with lower added noise $V_{3}=V_{1}$,
causing the entanglement properties of the output state to be the
same as in protocol 1.

In all schemes, if mode $B$ or $C$ is ignored, the CC never exceeds
$C^{\rm coh}$. For the scheme of Fig.~\ref{fig2},  $C^{2}$ and $C^{3}$ ($C^{1}$) exceed(s) $C^{\rm
coh}$ when $\bar{n}>0.36$ ($\bar{n}>0.44$). Note also that
$C^{3}\geq C^{2}$ because  protocol $3$ has lower noise than
protocol $2$. $C^{2}\geq C^{1}$ due to the symmetry of
protocol 2 with respect to the quadratures, which allows both
signals to be decoded with equal efficiency. In fact, $C^{3}$ even
exceeds $C^{\rm sq}$ for $\bar{n}>11.28$, which is a similar
result to the CC of controlled dense coding assisted by a
measurement on the collaborator's mode \cite{Jing_03}.

In conclusion, we have demonstrated experimentally that there are
fully separable and only globally non-classical three-mode states
that can lead to entanglement using a beam splitter. A similar
effect happens also in the qubit case, where the CNOT gate can
generate entanglement by acting on a part of a suitable
three-qubit fully separable state, whereas it leaves the output of
the operation separable \cite{Cubitt_03}. The local state may
appear unsuitable as a quantum resource. However, when being a
part of a larger correlated state, it can become a source of
tailored entanglement. This highlights the relevance of global
correlations in quantum technologies.


L. M. acknowledges the Project No. P205/12/0694 of GACR. N. K. is grateful for the support provided by the A. von Humboldt Foundation.
C. C. and N. K. acknowledge the support from the Scottish Universities Physics Alliance (SUPA) and the
Engineering and Physical Sciences Research Council (EPSRC). The project was supported within the framework of the
BMBF grant ``QuOReP'' and in the framework of the International Max Planck Partnership (IMPP) with Scottish Universities.

*C. C. and C. P. contributed equally to this work.

\end{document}